\documentclass[fleqn]{2017SCGE}

\usepackage{color}

\usepackage{mwe,tikz,ulem}
\usepackage{graphicx}
\usepackage[authoryear]{natbib}

\newcommand{\lad}{LAD\xspace}
\newcommand{\wfm}{WFM\xspace}

\newcommand{\lfa}{SFA\xspace}
\newcommand{\gpd}{PFA\xspace}

\newcommand{\rhosat}{$2.8\times 10^{14}$ g~cm$^{-3}$}
\newcommand{\rhos}{$\rho_\mathrm{sat}$}
\newcommand{\flux}{erg/cm$^2$/s}

\newcommand{\apj}{ApJ\xspace}
\newcommand{\apjl}{ApJ\xspace}
\newcommand{\apjs}{ApJ Supp.\xspace}
\newcommand{\mnras}{MNRAS\xspace}
\newcommand{\aap}{A\&A\xspace}

\newcommand{\nat}{Nature\xspace}
\newcommand{\araa}{ARAA\xspace}

\begin{document}

\ArticleType{Article}%??Article
\SpecialTopic{SPECIAL TOPIC: }%???????
\Year{2017}
\Month{January}
\Vol{60}
\No{1}
\DOI{10.1007/s11432-016-0037-0}
\ArtNo{000000}
\ReceiveDate{January 11, 2016}
\AcceptDate{April 6, 2016}

% title of the report
\title[Dense matter with eXTP]{Dense matter with eXTP}

\author[]{\parbox[t]{15cm}{
Anna L. Watts$^{1}$,
Wenfei Yu$^{2}$,
Juri Poutanen$^{3,4}$,
Shu Zhang$^{5}$,
Sudip Bhattacharyya$^{6}$,
Slavko Bogdanov$^{7}$,
Long Ji$^{8}$,
Alessandro Patruno$^{9}$,
Thomas E. Riley$^{1}$,
Pavel Bakala$^{10}$,
Altan Baykal$^{11}$,
Federico Bernardini$^{12,13}$,
Ignazio Bombaci$^{14,15}$,
Edward Brown$^{16}$,
Yuri Cavecchi$^{17,18}$,
Deepto Chakrabarty$^{19}$,
J\'er\^ome Chenevez$^{20}$,
Nathalie Degenaar$^{1}$,
Melania Del Santo$^{21}$,
Tiziana Di Salvo$^{22}$,
Victor Doroshenko$^{8}$,
Maurizio Falanga$^{23}$
Robert D. Ferdman$^{24}$,
Marco Feroci$^{25}$,
Angelo F. Gambino$^{22}$,
MingYu Ge$^{5}$,
Svenja K. Greif$^{26,27}$,
Sebastien Guillot$^{28}$,
Can Gungor$^{5}$,
Dieter H. Hartmann$^{29}$,
Kai Hebeler$^{26,27}$,
Alexander Heger$^{30}$,
Jeroen Homan$^{19}$,
Rosario Iaria$^{22}$,
Jean in 't Zand$^{31}$,
Oleg Kargaltsev$^{32}$,
Aleksi Kurkela$^{33,34}$,
Xiaoyu Lai$^{35}$,
Ang Li$^{36}$,
XiangDong Li$^{37}$,
Zhaosheng Li$^{38}$,
Manuel Linares$^{39}$
FangJun Lu$^{5}$,
Simin Mahmoodifar$^{40}$,
Mariano M{\'e}ndez$^{41}$,
M. Coleman Miller$^{42}$,
Sharon Morsink$^{43}$,
Joonas  N\"attil\"a$^{3,4}$,
Andrea Possenti$^{44}$,
Chanda Prescod-Weinstein$^{45}$,
JinLu Qu$^{5}$,
Alessandro Riggio$^{46}$,
Tuomo Salmi$^3$,
Andrea Sanna$^{46}$,
Andrea Santangelo$^{8,5}$,
Hendrik Schatz$^{47}$,
Achim Schwenk$^{26,27,48}$,
LiMing Song$^{5}$,
Eva {\v S}r\'amkov\'a$^{10}$,
Benjamin Stappers$^{49}$,
Holger Stiele$^{50}$
Tod Strohmayer$^{40}$,
Ingo Tews$^{51,47}$,
Laura Tolos$^{52,53,54}$
Gabriel T\"or\"ok$^{10}$,
David Tsang$^{18}$,
Martin Urbanec$^{10}$,
Andrea Vacchi$^{55,56}$,
RenXin Xu$^{57}$,
Ypeng Xu$^{5}$,
Silvia Zane$^{58}$,
Guobao Zhang$^{59}$,
ShuangNan Zhang$^{5}$,
Wenda Zhang$^{60}$,
ShiJie Zheng$^{5}$,
Xia Zhou$^{61}$
}}{}

\address[]{\parbox[t]{15cm}{
$^{1}$Anton Pannekoek Institute for Astronomy, University of Amsterdam, Science Park 904, 1098 XH Amsterdam, The Netherlands,
$^{2}$Shanghai Astronomical Observatory, Shanghai 200030, China,
$^{3}$Tuorla Observatory, Department of Physics and Astronomy, University of Turku, V\"ais\"al\"antie 20, FIN-21500 Piikki\"o, Finland,
$^{4}$Nordita, KTH Royal Institute of Technology and Stockholm University, Roslagstullsbacken 23, SE-10691 Stockholm, Sweden,
$^{5}$Institute of High Energy Physics, CAS, Beijing 100049, China,
$^{6}$Tata Institute of Fundamental Research, Mumbai 400005, India,
$^{7}$Columbia Astrophysics Laboratory, Columbia University, 550 West 120th Street, New York, NY 10027, USA,
$^{8}$Institut f\"ur Astronomie und Astrophysik T\"ubingen, Universit\"at T\"ubingen, Sand 1, D-72076 T\"ubingen, Germany,
$^{9}$Leiden Observatory, Leiden University, PO Box 9513, 2300 RA, Leiden, the Netherlands,
$^{10}$Research Center for Computational Physics and Data Processing,  Silesian University in Opava, Bezrucovo nam. 13, CZ-74601 Opava, Czech Republic,
$^{11}$Physics Department, Middle East Technical University, 06531 Ankara, Turkey,
$^{12}$INAF, Osservatorio Astronomico di Roma, Via Frascati 33, I-00078 Monteporzio Catone, Italy,
$^{13}$New York University Abu Dhabi, PO Box 129188, Abu Dhabi, UAE,
$^{14}$Dipartimento di Fisica Enrico Fermi, University of Pisa, I-56127 Pisa, Italy,
$^{15}$INFN Italian National Institute for Nuclear Physics, Pisa Section, 56127, Pisa, Italy,
$^{16}$Department of Physics and Astronomy, Michigan State University, East Lansing, MI 48824, USA,
$^{17}$Department of Astrophysical Sciences, Princeton University, Peyton Hall, Princeton, NJ 08544, USA,
$^{18}$Mathematical Sciences and STAG Research Centre, University of Southampton, Southampton, SO17 1BJ, UK,
$^{19}$MIT Kavli Institute for Astrophysics and Space Research, Cambridge, MA 02139, USA,
$^{20}$DTU Space, Technical University of Denmark, Elektrovej 327-328, 2800 Kgs Lyngby, Denmark,
$^{21}$INAF/IASF Palermo, via Ugo La Malfa 153, I-90146 - Palermo, Italy,
$^{22}$Universita degli Studi di Palermo, Dipartimento di Fisica e Chimica, via Archira 36, 90123 Palermo, Italy,
$^{23}$International Space Science Institute (ISSI), Hallerstrasse 6,CH-3012 Bern, Switzerland,
$^{24}$Faculty of Science, University of East Anglia, Norwich NR4 7TJ, UK,
$^{25}$INAF, Istituto di Astrofisica e Planetologie Spaziali, Via Fosso del Cavaliere 100, 00133 Roma, Italy,
$^{26}$Institut f\"ur Kernphysik, Technische Universit\"at Darmstadt, D-64289 Darmstadt, Germany,
$^{27}$ExtreMe Matter Institute EMMI, GSI Helmholtzzentrum f\"ur Schwerionenforschung GmbH, D-64291 Darmstadt, Germany,
$^{28}$Instituto de Astrof\'isica, Pontificia Universidad Cat\'olica de Chile, Av. Vicu\~na Mackenna 4860, 782-0436 Macul, Santiago, Chile,
$^{29}$Department of Physics \& Astronomy, Kinard Lab of Physics, Clemson University, Clemson, SC 29634-0978, USA,
$^{30}$School of Physics and Astronomy, Monash University, Clayton VIC 3800, Australia,
$^{31}$SRON Netherlands Institute for Space Research, Sorbonnelaan 2, 3584 CA Utrecht, The Netherlands,
$^{32}$Department of Physics, The George Washington University, 725 21st St. NW, Washington, DC 20052, USA,
$^{33}$Theoretical Physics Department, CERN, Geneva, Switzerland,
$^{34}$Faculty of Science and Technology, University of Stavanger, 4036 Stavanger, Norway,
$^{35}$School of Physics and Mechanical \& Electrical Engineering, Hubei University of Education, Wuhan 430205, China,
$^{36}$Department of Astronomy, Xiamen University (Haiyun Campus) Xiamen, Fujian 361005, China,
$^{37}$School of Astronomy and Space Science, Nanjing University, Nanjing 210023, China,
$^{38}$Department of Physics, Xiangtan University, Xiangtan 411105, China,
$^{39}$Departament de F{\'i}sica, EEBE, Universitat Polit{\`e}cnica de Catalunya, c/ Eduard Maristany 10, 08019 Barcelona, Spain
$^{40}$NASA's Goddard Space Flight Center, Greenbelt, MD 20771, USA,
$^{41}$Kapteyn Astronomical Institute, University of Groningen, Postbus 800, NL-9700 AV Groningen, the Netherlands,
$^{42}$Department of Astronomy and Joint Space-Science Institute, University of Maryland, College Park, MD 20742-2421, USA,
$^{43}$Department of Physics, University of Alberta, Edmonton T6G 2E1, Alberta, Canada,
$^{44}$INAF, Osservatorio Astronomico di Cagliari, Via della Scienza 5, I-09047 Selargius, Italy,
$^{45}$Department of Physics, University of Washington, Seattle, Washington 98195-1560, USA,
$^{46}$Dipartimento di Fisica, Universita degli Studi di Cagliari, SP Monserrato-Sestu km 0.7, 09042 Monserrato, Italy,
$^{47}$JINA Center for the Evolution of the Elements, National Superconducting Cyclotron Laboratory, Michigan State University, East Lansing, MI 48824, USA,
$^{48}$Max-Planck-Institut f{\"u}r Kernphysik, Saupfercheckweg 1, D-69117 Heidelberg, Germany,
$^{49}$Jodrell Bank Centre for Astrophysics, School of Physics and Astronomy, The University of Manchester, Manchester M13 9PL, UK,
$^{50}$Institute for Astronomy, National Tsing Hua University, Guangfu Road 101, Sect. 2, 30013 Hsinchu, Taiwan (R.O.C.),
$^{51}$Institute for Nuclear Theory, University of Washington, Seattle, WA 98195, USA,
$^{52}$Institute of Space Sciences, IEEC-CSIC, Carrer Can Magrans s/n, 08193, Barcelona, Spain,
$^{53}$Institut f{\"u}r Theoretische Physik, Goethe-Universit{\"a}t Frankfurt, Max-von-Laue-Str. 1, 60438 Frankfurt am Main, Germany,
$^{54}$Frankfurt Institute for Advanced Studies, Ruth-Moufang-Str. 1, 60438 Frankfurt am Main, Germany,
$^{55}$Department of Mathematics, Computer Science and Physics, University of Udine  33100 Udine, Italy,
$^{56}$INFN Italian National Institute for Nuclear Physics, Trieste Section, 34149 Trieste, Italy,
$^{57}$School of Physics and State Key Laboratory of Nuclear Physics and Technology, Peking University, Beijing 100871, China,
$^{58}$Mullard Space Science Laboratory, University College London, Holmbury St. Mary, Dorking, Surrey RH5 6NT, UK,
$^{59}$Yunnan Observatory, The Chinese Academy of Sciences, Kunming 650011, China 
$^{60}$Astronomical Institute of the Academy of Sciences, Bo{\v c}n\'i II 1401, CZ-14100 Praha 4, Czech Republic,
$^{61}$Xinjiang Astronomical Observatory, Chinese Academy of Sciences, Urumqi 830011, China.
}}{}

\AuthorMark{Watts A L, Yu W, Poutanen J, Zhang S, et al}
\AuthorCitation{Watts A L, Yu W, Poutanen J, Zhang S, et al}
%\titlecitation{Dense matter with eXTP}

\abstract{In this White Paper we present the potential of the \textit{Enhanced X-ray Timing and Polarimetry} (\textit{eXTP}) mission for determining the nature of \textit{dense matter}; neutron star cores host an extreme density regime which cannot be replicated in a terrestrial laboratory.  The tightest statistical constraints on the dense matter equation of state will come from pulse profile modelling of accretion-powered pulsars, burst oscillation sources, and rotation-powered pulsars.  Additional constraints will derive from spin measurements, burst spectra, and properties of the accretion flows in the vicinity of the neutron star. Under development by an international Consortium led by the Institute of High Energy Physics of the
Chinese Academy of Science, the eXTP mission is expected to be launched in the mid 2020s.}

\keywords{stars: neutron; X-rays: stars; dense matter; equation of state}

\PACS{26.60.Kp, 95.55.Ka, 97.60.Jd}

\maketitle

\begin{multicols}{2}

\section{Introduction}

The  \textit{enhanced X-ray Timing and Polarimetry mission} (\textit{eXTP}) is a mission concept proposed by a consortium led by the Institute of
  High-Energy Physics of the Chinese Academy of Sciences, envisaged for a launch in the mid 2020s.  \textit{eXTP} would carry 4 instrument packages for
  the 0.5--30 keV bandpass, its primary purpose being to study conditions of extreme density (this paper), gravity \citep{WP_SG} and magnetism \citep{WP_SM} in and around compact objects in the Universe.  It would also be a powerful observatory for a wider range of astrophysical phenomena since it combines high throughput, good spectral and timing resolution, polarimetric capability and wide sky coverage \citep{WP_OS}.  

A detailed description of eXTP's instrumentation can be found in \citet{WPinstrumentation}, but we summarize briefly here.  The scientific payload of eXTP consists of the Spectroscopic Focusing Array (\lfa), the
Polarimetry Focusing array (\gpd), the Large Area Detector (LAD), and the Wide Field Monitor (\wfm).  
The \lfa is an array of nine identical X-ray telescopes covering the energy range 0.5--10~keV with a spectral resolution of better than 180 eV (full width at half maximum, FWHM) at 6 keV, and featuring a total
effective area from $\sim$0.7~m$^2$ at 2~keV to $\sim 0.5$~m$^2$ at
6~keV.   The \lfa angular resolution is required to be less than 1 arcmin (HPD).
In the current baseline, the \lfa focal plane detectors
are silicon-drift detectors (SDDs), that combine CCD-like spectral
resolutions with very small dead times, and therefore are excellently
suited for studies of the brightest cosmic X-ray sources at the
smallest time scales.  The \gpd consists of four identical X-ray telescopes that are
  sensitive between 2 and 8 keV with a spectral resolution of 1.1 keV
  at 6 keV (FWHM), have an angular resolution better than $\sim30$~arcsec (HPD) and a
total effective area of $\sim 900$~cm$^2$ at 2~keV (including the
detector efficiency). The \gpd features Gas Pixel Detectors (GPDs) to
allow polarization measurements in the X-rays. It reaches a minimum
detectable polarization (MDP) of 5\% in 100~ks for a source with the Crab-like spectrum
of flux $3\times10^{-11}$~erg~s$^{-1}$~cm$^{-2}$ (i.e. about 1 milliCrab). The \lad has a very large effective area of $\sim 3.4$~m$^2$ at 8~keV,
obtained with non-imaging SDDs, active between 2 and 30~keV with a
spectral resolution of about 260 eV and collimated to a field of view
of 1$^\circ$ FWHM. The \lad and the \lfa together reach an
unprecedented total effective area of more than 4 ~m$^{2}$. The science payload is completed by the \wfm, consisting of 6
coded-mask cameras covering about 4~sr of the sky at at an expected sensitivity of 2.1 mCrab for an exposure time of 50 ks in the 2 to 50~keV energy range,
and for a typical sensitivity of 0.2~mCrab combining 1~yr of
observations outside the Galactic plane. The instrument will feature
an angular resolution of a few arcminutes and will be endowed with an
energy resolution at 6 keV of about 300 eV (FWHM).

The nature of matter under conditions of extreme density and stability, found only in the cores of neutron stars (NSs), remains an open question.  eXTP's capabilities will allow us to statistically infer global properties of NSs (such as their mass and radius) to within a few percent.  This information can be used to statistically infer the equation of state of the matter in the NS interior, and the nature of the forces between fundamental particles under such extreme conditions.  This White Paper outlines the current state of our understanding of dense matter physics, the techniques that eXTP will exploit, and the advances that we expect.

\section{The nature of dense matter}

\begin{figure*}[!hb]
\includegraphics[width=1.0\textwidth]{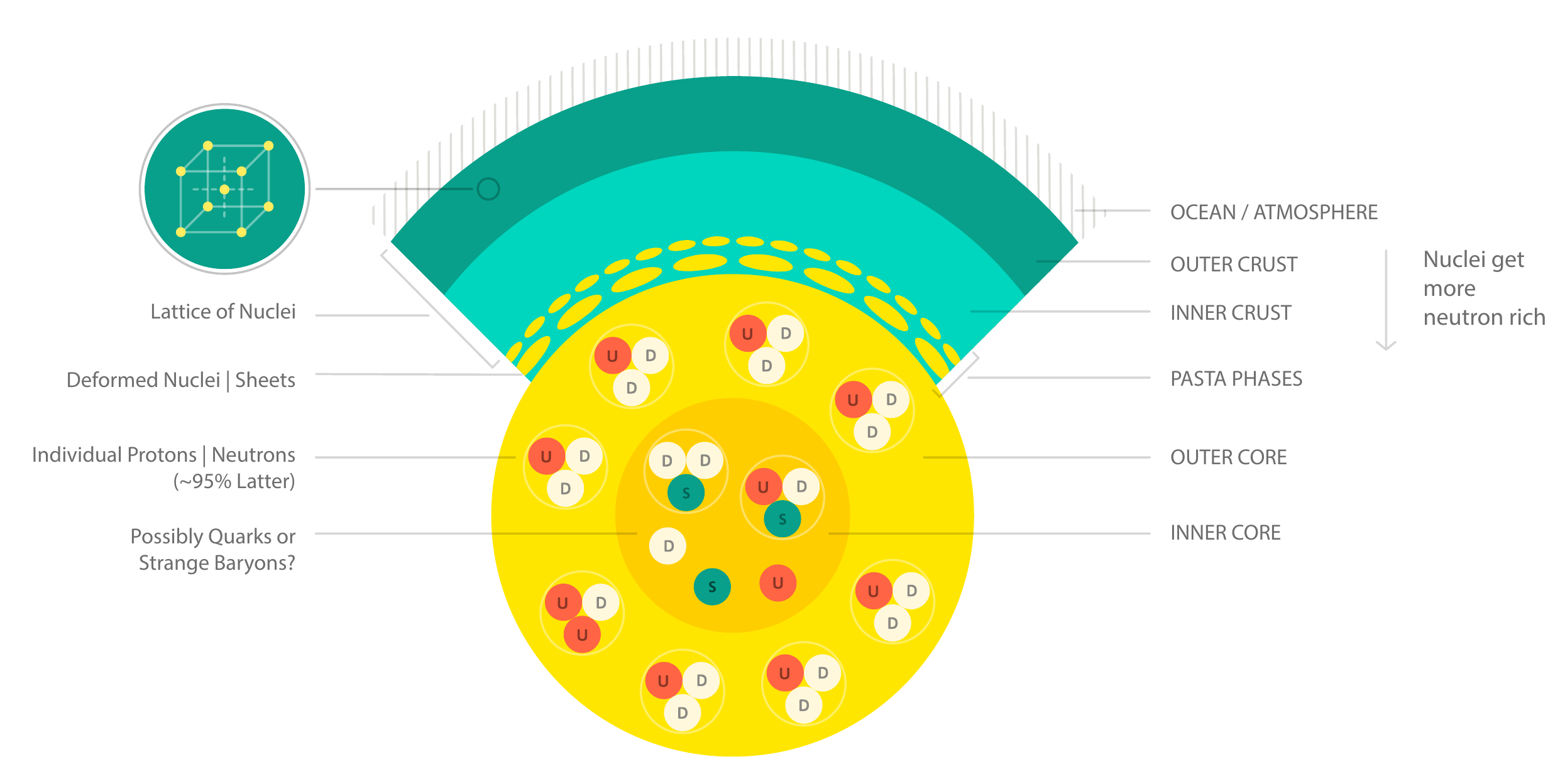}
\caption{Schematic structure of a NS (not to scale). The outer layer is a solid ionic crust supported by electron degeneracy pressure. Neutrons begin to leak out of ions (nuclei) at densities $\sim 4\times 10^{11}$ g~cm$^{-3}$ (the neutron drip density, which separates inner from outer crust), where neutron degeneracy also starts to play a role.  At the very base of the crust, nuclei may become very deformed (the pasta phase).  At densities $\sim 2 \times 10^{14}$ g~cm$^{-3}$ , the nuclei dissolve completely and this marks the crust-core boundary. In the core, densities could reach up to ten times \rhos $\sim$ \rhosat (\rhos being the density in normal atomic nuclei). The matter in the core is highly neutron-rich, and the inner core may contain stable states of strange matter in deconfined quark or baryonic form. See also Figure 1 of \citet{Watts16}.}
\label{fig:nscut}
\end{figure*}

One of the overarching goals of modern physics is to understand the nature of the fundamental interactions. Here we focus on the strong interaction, which controls the properties of both atomic nuclei and NSs, where gravity compresses the material in the core of the star to extreme nuclear densities (Figure \ref{fig:nscut}).  NSs are remarkable natural laboratories that allow us to investigate the constituents of matter and their fundamental interactions under conditions that cannot be reproduced in any terrestrial laboratory, and to explore the phase diagram of quantum chromodynamics (QCD) in a region which is presently inaccessible to numerical calculations \citep{Fukushima11}. 

\begin{figure*}[!ht]
\centering
\includegraphics[width=1.0\textwidth]{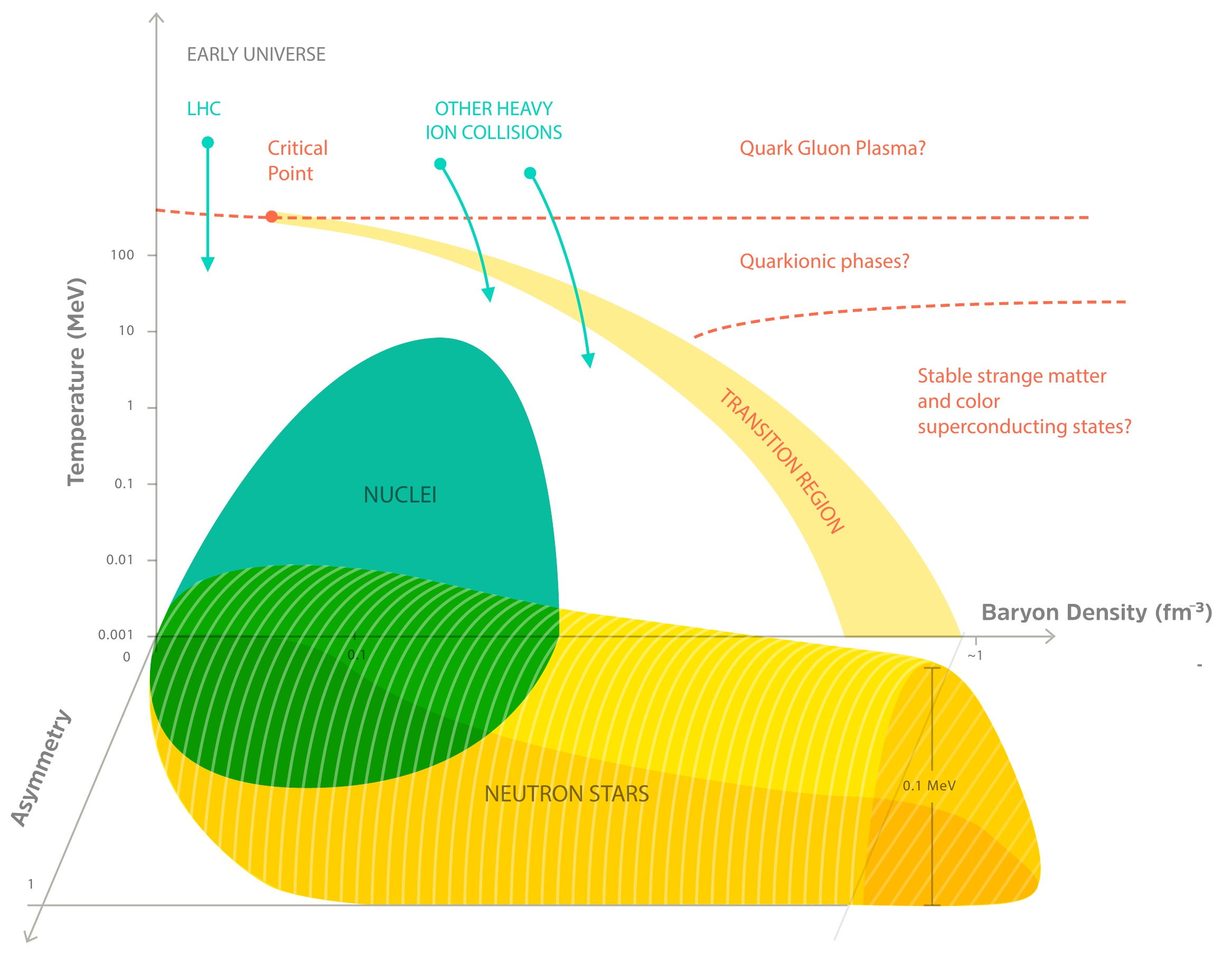}
\caption{Hypothetical states of matter accessed by NSs and current or planned laboratory experiments, in the parameter space of temperature, baryon number density and nuclear asymmetry $\alpha = 1-2Y$ where $Y$ is the hadronic charge fraction ($\alpha = 0$ for matter with equal numbers of neutrons and protons, and $\alpha = 1$ for pure neutron matter). NSs access unique states of matter that are either difficult to create in the laboratory - such as nuclear superfluids and perhaps strange matter states like hyperons - or cannot be created, such as deconfined quarks and the color superconductor phase. For simplicity, the transition region is shown only in projection on the density-temperature axis.  Figure adapted from Fig.~2 of \citet{Watts15}; for an alternative visualisation in the parameter space of temperature and baryon chemical potential, see Fig.~2 of \citet{Watts16}. }
\label{fig:trho}
\end{figure*}

The quest to test the state of matter under the most extreme conditions and to determine the equation of state (EOS) encompasses both laboratory experiments and astronomical observations of stars (Fig.~\ref{fig:trho}).  Heavy-ion collision experiments currently going on at the Relativistic Heavy Ion Collider (RHIC) at Brookhaven and at the Large Hadron Collider (LHC) at CERN can probe the high temperature and 
low density region of the strong interacting matter phase diagram.   
The next generation of heavy-ion colliders such as the Facility for Antiproton and Ion Research (FAIR) at GSI 
in Darmstadt, and the Nuclotron-based Ion Collider fAcility (NICA) at JINR in Dubna will be able 
to probe high temperature and dense matter (up to $\sim 4$ \rhos, see Figure \ref{fig:nscut}) and to search for the possible existence of a critical endpoint of a first-order quark deconfinement phase transition.  Laboratory EOS constraints through heavy ion collisions will also be pursued at rare isotope facilities such as RIKEN/RIBF and FRIB (where the collisions will have less energy but more neutron richness).  

Neutron stars, by contrast, access a unique region of the QCD phase diagram at low temperature 
($T \ll 1$~MeV after a few minutes from the NS birth) and high density (up to $\sim 10$ \rhos) 
which cannot be explored in the laboratory.  In the simplest picture the core of a NS is modeled as a uniform charge-neutral fluid of neutrons $n$, protons $p$, electrons $e^-$ and muons $\mu^-$ in equilibrium with respect to the weak interaction  ($\beta$-stable nuclear matter).  Even in this simplified picture, the determination of the EOS from the underlying nuclear interactions is a formidable theoretical problem. One has to calculate the EOS under extreme conditions of 
high density and high neutron-proton asymmetry \citep[see for example][and Figure \ref{fig:trho}]{Hebeler15}, in a regime where the 
properties of nuclear forces are poorly constrained by nuclear data and experiments.  

Due to the large central densities additional constituents, such as hyperons (\citealt{Glendenning85}, \citealt{Chatterjee16}) or a quark deconfined phase of matter (see for example \citealt{Glendenning96} and \citealt{Bombaci2016}) may also form.  The reason for hyperon formation is simple: the stellar constituents $npe^-\mu^-$  are different species of fermions, so due to the Pauli principle their Fermi energies (chemical potentials) are very rapidly increasing functions of the density.   Above some threshold density 
($\sim$ 2-4 \rhos) it is energetically favorable to form hyperons via the strangeness-changing weak interaction.  This means that there may be different types of compact stars - nucleonic, hyperonic, hybrid or quark - the latter two containing deconfined up-down-strange quark matter in their cores.
 
Various superfluid states produced through Cooper pairing (caused by an attractive component of 
the baryon-baryon interaction) are also expected.  For example, a neutron superfluid (due to neutron-neutron pairing in the $^1S_0$ channel)  is expected in the NS inner crust. Many possible color superconducting phases of quark matter  
are also expected \citep{Alford08} in quark deconfined matter.  Matter may also be characterized by the formation of different crystalline structures \citep{Anglani14,Buballa15}.  These superfluid, color superconducting and crystalline phases of matter are of crucial importance for modeling NS cooling and pulsar glitches.  

\begin{figure*}[!ht]
\includegraphics[width=0.48\textwidth]{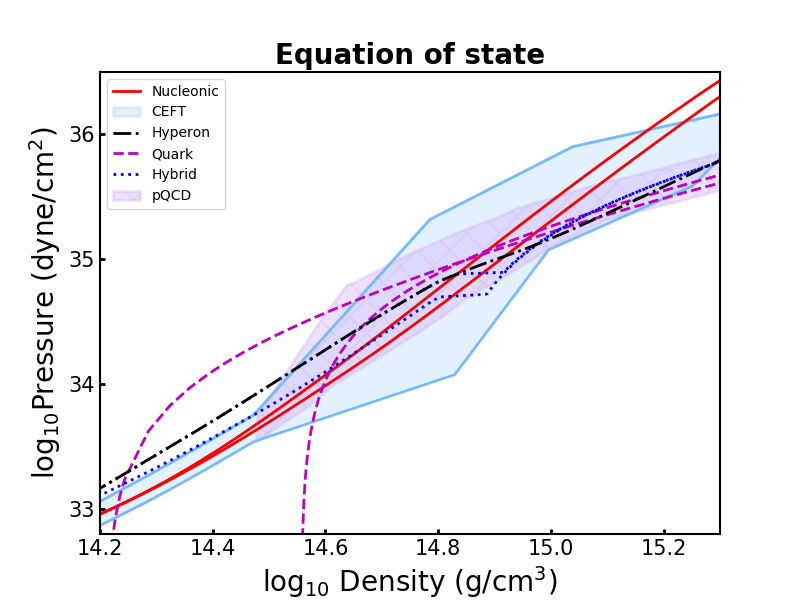}
\includegraphics[width=0.48\textwidth]{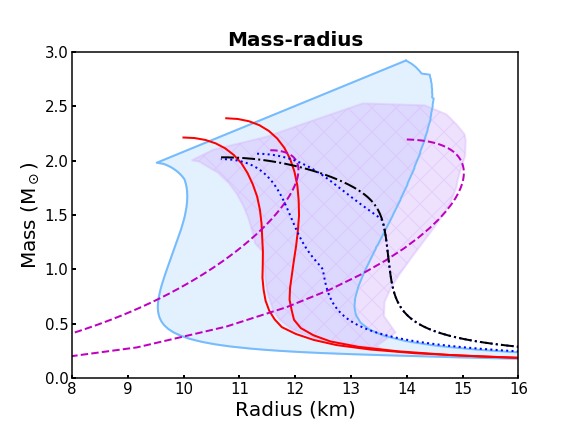}
\caption{The pressure--density relation (EOS, left) and the corresponding M-R relation (right) for some example models with different microphysics.  Nucleonic (neutrons, protons):  models AP3 and AP4 from \citet{Akmal97}, also used in \citet{Lattimer01}. Quark (u, d, s quarks): models from \citet{Li16} and \citet{Bhattacharyya16}.  Hybrid (inner core of uds quarks, outer core of nucleonic matter): models from \citet{Zdunik13}.  Hyperon (inner core of hyperons, outer core of nucleonic matter): Model from  \citet{Bednarek12}.  CEFT:  range of nucleonic EOS based on Chiral Effective Field Theory (CEFT) from \citet{Hebeler13}.  pQCD: range of nucleonic EOS from \citet{Kurkela14} that interpolate from CEFT at low densities and match to perturbative QCD (pQCD) calculations at higher densities than shown in this figure. All of the EOS shown are compatible with the existence of $\sim 2$ M$_\odot$ NSs.}
\label{fig:eostomr}
\end{figure*}

Connecting NS parameters to strong interaction physics can be done because the forces between the nuclear particles set the stiffness of NS matter \citep{Lattimer16}.  This is encoded in the EOS, the thermodynamical relation between pressure, energy density and temperature.   The EOS of dense matter is a basic ingredient for modeling various astrophysical phenomena related to NSs, including core-collapse supernovae and binary neutron star mergers (note that for most neutron star scenarios -- except immediately after formation or merger -- we can consider the temperature to be effectively zero).   The EOS and NS rotation rate set the gravitational mass M and equatorial radius R via the stellar structure equations.  By measuring and then inverting the M-R relation, we can thus recover the EOS \citep[][and Figure \ref{fig:eostomr}]{Lindblom92,Ozel09,Riley18}. To distinguish the models shown in Figure \ref{fig:eostomr}, one needs to measure M and R to precisions of a few percent, for multiple sources with different masses.

Most efforts to date to measure the M-R relation have involved modelling the spectra of thermonuclear X-ray bursts and quiescent low-mass X-ray binaries (see for example \citealt{Suleimanov11}, \citealt{Ozel13}, \citealt{Steiner13}, \citealt{Guillot14}, \citealt{Nattila16}, \citealt{Nattila17}, \citealt{Steiner17}). The constraints obtained so far are weak.  The technique also suffers from systematic errors of at least 10\% in absolute flux calibration, and uncertainties in atmospheric composition, residual accretion, non-uniform emission, distance, and identification of photospheric touchdown point for bright bursts that exhibit Photospheric Radius Expansion (PRE) (see the discussions in \citealt{Miller13a}, \citealt{Heinke14}, \citealt{Poutanen14}).  The planned ESA L-class mission Athena has the right energy band to exploit this technique \citep{Motch13}: however the systematic uncertainties will remain.  The X-ray timing instrument NICER (the Neutron Star Interior Composition Explorer, see \citealt{Arzoumanian14}), which was installed on the International Space Station in 2017 and which will instead use the pulse-profile modelling technique, is discussed in more detail in Section \ref{rpp}. 

Constraints have also been derived via radio pulsar timing, where the masses of NSs in compact binaries can be measured very precisely: high mass stars yield the strongest EOS constraints.  However even the discovery of pulsars with masses $\approx 2 $M$_\odot$ (\citealt{Demorest10}, \citealt{Antoniadis13}, \citealt{Fonseca16}) has left a broad range of EOS viable, producing radii ranging from 10--14 km for a typical 1.4 M$_\odot$ NS \citep{Hebeler13}.  The next generation of radio telescopes (the Square Kilometer Array and its precursors) will deliver improved mass measurements.  Precision radius measurements, however, will be more challenging: there is only one system, the Double Pulsar, for which we expect a radius measurement with $\sim$ 5-10\% accuracy (via its moment of inertia) within the next 20 years (\citealt{Lattimer05}, \citealt{Kramer09}, \citealt{Watts15}).

\begin{figure*}[!ht]
\includegraphics[width=0.9\textwidth]{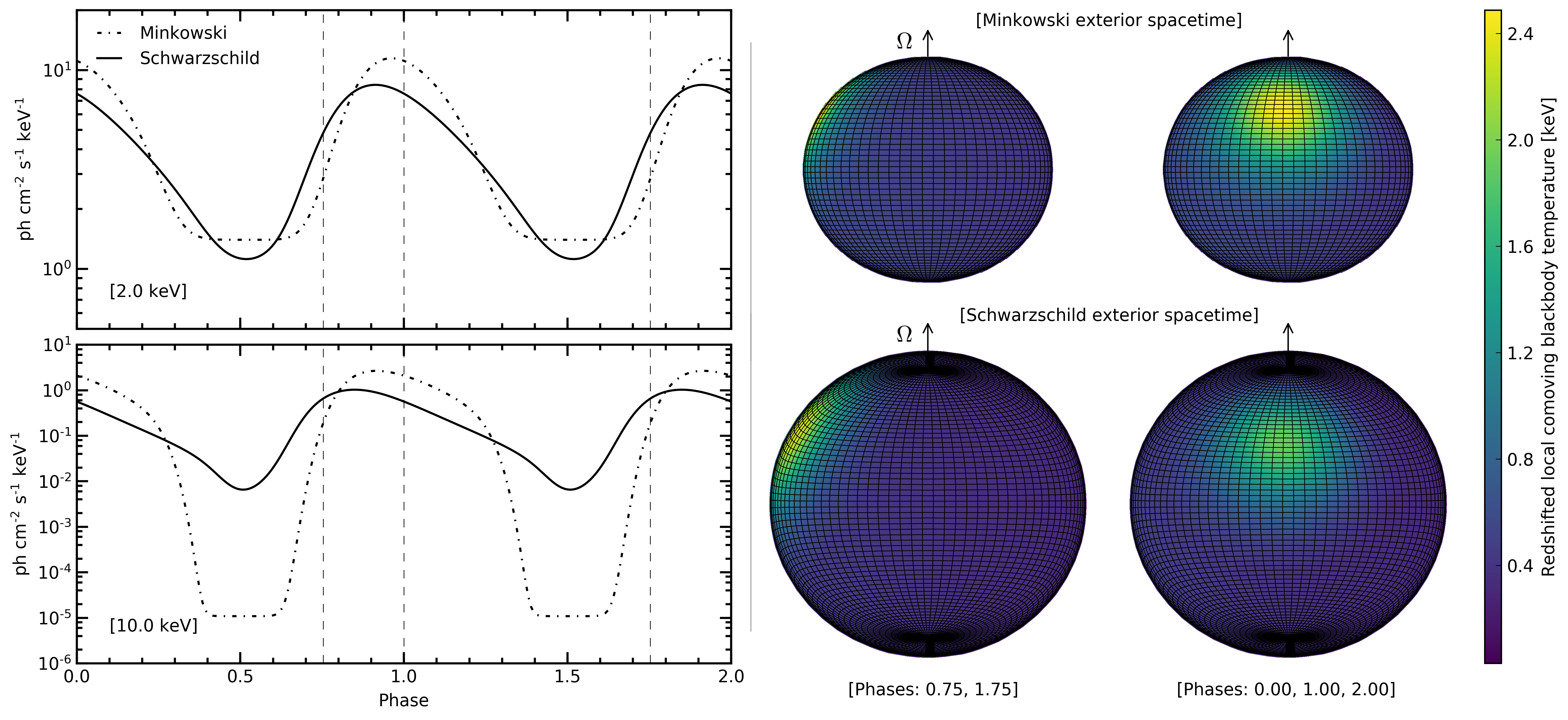}
\caption{Stellar rotation modulates emission from a hot region (hotspot), generating an X-ray pulsation. Relativistic effects encode information about M and R in the normalisation and harmonic content of the pulse profile. These effects are key observables exploited by the pulse profile modelling technique, and include Doppler boosting and gravitational redshifting, time-delays, and light bending (which renders the far side of the star partially visible). The Figure illustrates these effects for a rapidly spinning, oblate star.  We compare pulse profiles generated by a photosphere embedded in a Minkowski exterior spacetime to those generated by a photosphere embedded in a Schwarzschild exterior spacetime (see text).  For the purpose of illustration, we use: a gravitational mass of 1.8 M$_\odot$; an equatorial coordinate radius of 14 km; a coordinate spin frequency of 600 Hz; and a distance of 1~kpc. The observer is in the equatorial plane. The local photospheric radiation field is completely specified by the local comoving blackbody temperature. The temperature field is non-evolving in the corotating reference frame, and is constituted by a hotspot of angular radius of 60$^\circ$, centred at a colatitude of 60$^\circ$. Its temperature falls smoothly from 2.5 keV at the centre to 0.5 keV at the boundary, where the latter is the temperature everywhere outside the spot boundary.  Left: Monochromatic profiles at two energies (2 and 10 keV).  Right: The resolved stellar photosphere at two rotational phases. The colour corresponds to the redshifted temperature on a distant image-plane. }
\label{fig:gr1}
\end{figure*}

The gravitational wave telescopes Advanced LIGO \citep{LIGO15} and Advanced VIRGO \citep{Acernese15}, have now made the first direct detection of a binary NS merger \citep{Abbott17}.  Gravitational waves from the late inspirals of binary NSs are sensitive to the EOS, with departures from the point particle waveform due to tidal deformation encoding information about the EOS \citep{Read09}.  The statistical constraints from the first detection are comparable to and in agreement with those obtained from X-ray spectral fitting.  In the event of a very high signal to noise event, Advanced LIGO/VIRGO may be able to constrain R to $\sim 10$\% (\citealt{Read13}, \citealt{Hotokezaka16}).  More realistic estimates indicate a few tens of detections are likely to be required to reach this level of accuracy (\citealt{Delpozzo13}, \citealt{Agathos15}, \citealt{Lackey15}, \citealt{Chatziioannou15}).  There may also be systematic errors of comparable size due to approximations made or higher-order terms neglected in the templates \citep{Favata14,Lackey15}.  The coalescence can also excite post-ringdown oscillations in the supermassive NS remnant that may exist very briefly before collapse to a black hole.  These oscillations are sensitive to the finite temperature EOS \citep{Bauswein12,Bauswein14,Takami14}, but detection will be difficult because there are no complete waveform models for the pre- and post-merger signal \citep{Clark16}.   The eventual detection of NS-black hole binary mergers may also yield EOS constraints (see for example \citealt{Lackey14}). See \citet{WP_OS} for other aspects of compact object merger astrophysics where eXTP can provide information on electromagnetic counterparts.  

The large area and spectral-timing-polarimetric capabilities of eXTP open up new techniques and different sources to constrain the dense matter EOS, which should allow us to measure M and R to within a few percent.  In the Sections that follow, we outline the various techniques that eXTP will use to measure the dense matter EOS, and explore its expected performance in more detail.

\section{Pulse profile modelling}
\label{ppm}
\subsection{Basic principles of pulse profile modelling }
\label{ppmb}

Pulse profile modelling exploits localised, radiatively intense regions (hereafter `hotspots', specific examples of which will be discussed in subsequent sections) that can develop on the NS.  As the star rotates, a hotspot generates an observable pulsation in X-rays. Prior to observation, the photons propagate through the curved exterior spacetime of the spinning compact star. Extensive work on propagation of electromagnetic radiation through such spacetimes has now quantified fully the relativistic effects on the photons, and thus on the pulse profile (\citealt{Pechenick83}, \citealt{Miller98}, \citealt{Poutanen03}, \citealt{PoutanenBeloborodov06}, \citealt{Cadeau07}, \citealt{Morsink07}, \citealt{Baubock13}, \citealt{Psaltis14a}, \citealt{NattilaPihajoki17}); the simulations in Figure \ref{fig:gr1} illustrate such observables, using a realistic Schwarzschild exterior spacetime.

\begin{figure*}[!ht]
\includegraphics[width=0.9\textwidth]{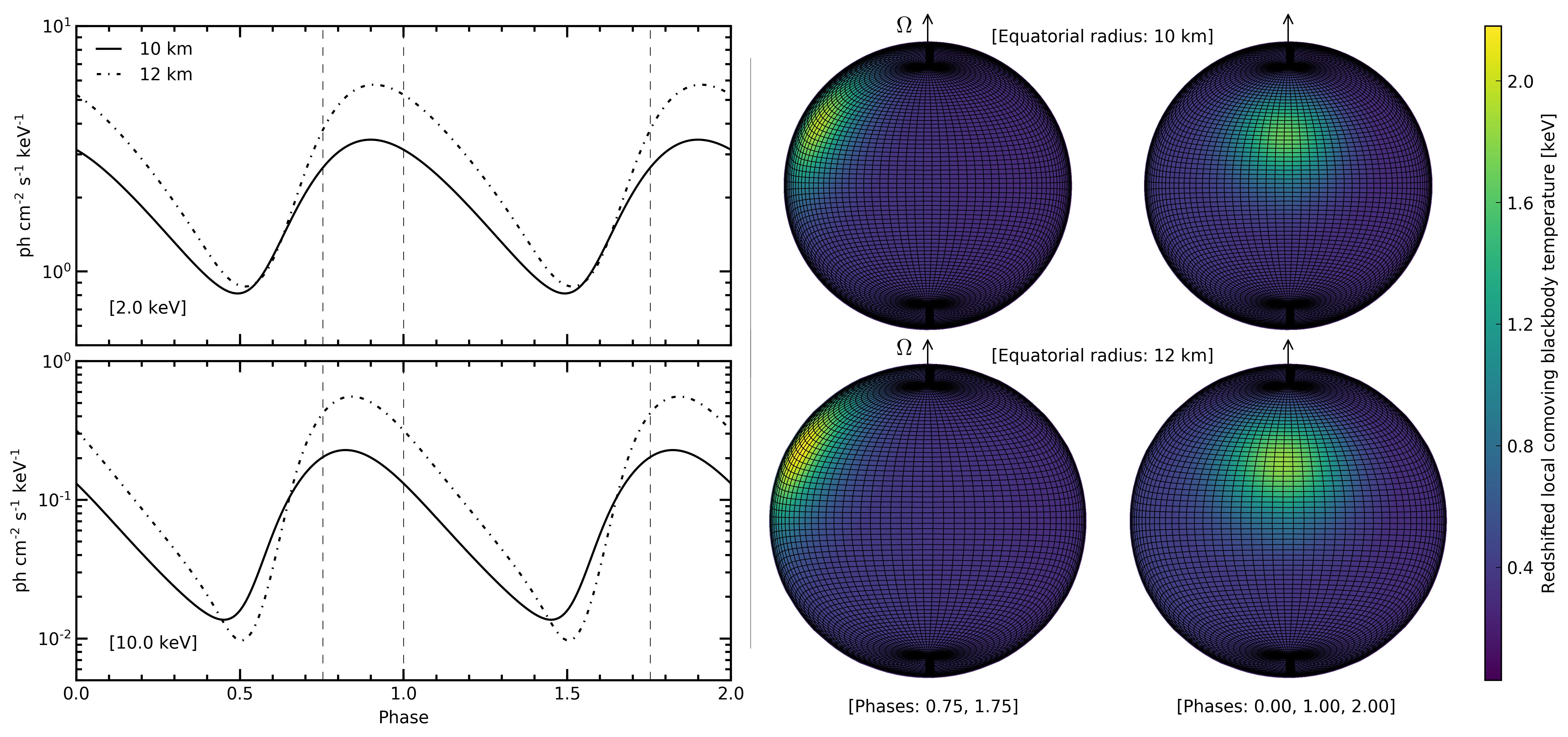}
\caption{We illustrate the response of monochromatic pulse profiles to variation of the (circumferential) equatorial radius, whilst all other parameters are fixed at the values implemented in Figure \ref{fig:gr1}. We use the realistic exterior spacetime described in Figure \ref{fig:gr1}. The two synthetic stars shown are of equal gravitational mass and spin frequency, but have equatorial radii of 10 km and 12 km; these stars require distinct EOS models to exist.  The pulse profiles are clearly sensitive to the equatorial radius, and it is the need to detect such differences that drives the design requirements for eXTP.}
\label{fig:gr2}
\end{figure*}

Strictly, the Schwarzschild exterior spacetime is exact only for spherically symmetric (stress-isotropic, non-rotating) stars.  However the mathematical structure of both the interior and exterior spacetime of a spinning NS is well understood in general relativistic gravity; high-accuracy spacetimes for rapidly spinning NSs can be computed numerically, albeit expensively (for a review, see \citealt{Stergioulas03}). For families of EOS, there exist (numerically computed) approximate universalities relating first-order (and higher) spacetime structure to the lowest-order properties -- specifically, the mass monopole moment, the (circumferential) equatorial radius, and the spin frequency (see for example the review of \citealt{Yagi16}).  In order to simulate observable radiation for the application of statistical inference, various approximations are employed which demonstrably reduce computation time. Given universal relations, one typically embeds an oblate surface -- from which radiation emanates -- in an ambient (exterior) spacetime, and either: (i) exploits spherical symmetry of the exterior (Schwarzschild) solution (see for example \citealt{Morsink07}); or (ii) permits axisymmetry, but neglects structure beyond second-order in a metric expansion in terms of a natural variable (see for example \citealt{Baubock12} or \citealt{NattilaPihajoki17} and references therein). The accuracy of these approximations are well understood (see the discussion in \citealt{Watts16}); embedding an oblate star in an ambient Schwarzschild spacetime introduces negligible systematic errors in the best-fit masses and radii at spin rates typical for observed millisecond pulsars.  We expect the statistical uncertainty incurred due to noise in eXTP observations to dominate systematic biases which would arise from low-order exterior spacetime approximation.  In practice this should be proven for each relevant generative model via blind parameter estimation studies, given synthetic data generated using a higher-order exterior spacetime. Nevertheless, in the coming years, algorithmic advances which improve both numerical likelihood evaluation speeds (via, e.g., extensive GPU exploitation) and Bayesian posterior sampling efficiencies may permit us to condition on generative models using higher-order exterior spacetimes.

\begin{figure*}
\includegraphics[width=0.9\textwidth]{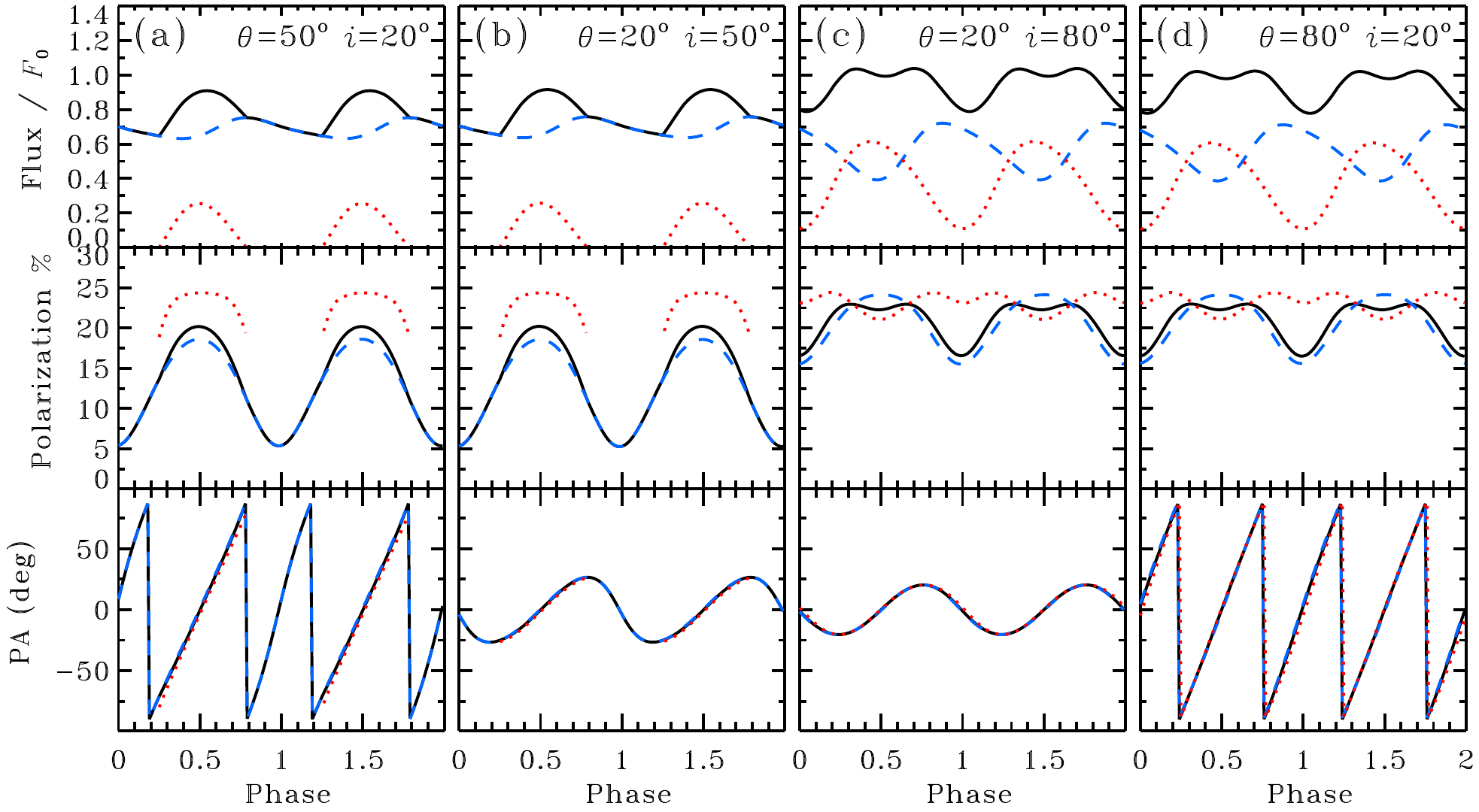}
\caption{The pulse profile as well as the phase-dependence of the PD and PA. 
The black solid line gives the contribution of two antipodal spots, while the blue dashed and red dotted lines correspond 
to the contribution of the primary and secondary spot separately. The pulse profile and PD are degenerate to exchanging $i$ and $\theta$,
while the PA shows dramatically different behaviour allowing both angles to be obtained  \citep[adapted from][]{Viironen04}. }
\label{fig:amps_polarization}
\end{figure*}

We now describe in a simplified manner how relativistic effects encode information on M and R. General relativistic (hereafter GR) light-bending, which is highly sensitive to compactness M/R in the near vicinity of the NS, directly affects both the amplitude of the pulsations and photon time-delays from distinct points on the NS surface. Gravitational redshifting of photons is also entirely dependent on the compactness, and manifests principally in the energy-dependent normalisation of the pulse profile.  Relativistic beaming introduces asymmetry (harmonic content) in the pulse profile; locally, beaming depends on the projected velocity of the (relativistically moving) hotspot along a light-ray connecting a point on the local NS surface to the observer. The functional form of the local speed contains R and the (asymptotic) spin frequency of the uniformly rotating star; these two parameters are degenerate with respect to influence on local beaming. However, the spin frequency can be accurately measured from the observed pulse frequency, thus breaking this degeneracy and increasing the statistical constraining power on R.  Figure \ref{fig:gr2} demonstrates the sensitivity of the observable to changing R alone, with all other model parameters fixed. The beaming is also sensitive to local time dilation at the NS surface, which is in turn sensitive to the compactness (M/R). The pulse profile model enters in a generative model for telescope photon data, and thus — via existing fitting algorithms — yields a statistical constraint on M and R.

Naturally, there are additional model parameters affecting the pulse profile, which must be properly marginalised over when statistically inferring M and R. These include the specific details of the photospheric comoving radiation field (as a function of surface coordinates, emission direction, and energy), and \textit{a priori} unknown geometrical factors (the hotspot size, shape, and colatitude $\theta$; observer inclination $i$), and emission from the rest of the star and disk, which may also exhibit pulsations \citep{Poutanen08}. However, the resulting degeneracies can be broken, allowing successful recovery of M and R (\citealt{Lo13}, \citealt{Psaltis14b}, \citealt{Miller15}, \citealt{Stevens16}).  Knowledge of the geometrical factors (enabled by the polarimetry capabilities of eXTP) further improves statistical constraining power via degeneracy breaking: M, R, and the nuisance parameters all enter in generative models for additional observable quantities.

Radiation emitted by hotspots is expected to be linearly polarised because the opacity is dominated by electron scattering \citep{Viironen04}. 
Both the observed polarization degree (PD) and polarization angle (PA) change with the  rotational phase $\phi$ 
following variations of the angle between the spot normal and the line-of-sight and of the position angle of the projection of the hotspot normal on the sky (see Fig.~\ref{fig:amps_polarization}). 
The variation of PA $\chi$  can be well described by the rotating vector model \citep{Radhakrishnan69}: 
\begin{equation} \label{eq:PA_RVM}
\tan\chi =-\frac{\sin \theta\ \sin \phi}
{\sin i\ \cos \theta  - \cos i\ \sin \theta\  \cos \phi }.
\end{equation}
This formula can be corrected for rapid rotation \citep{Ferguson73,Ferguson76} and gravitational light bending, but these effects are non-negligible only for spins in excess of 500 Hz \citep{Viironen04}. The phase-dependence of the PA allows us to constrain both angles $i$ and $\theta$.

\begin{figure*}[!ht]
\includegraphics[width=0.9\textwidth]{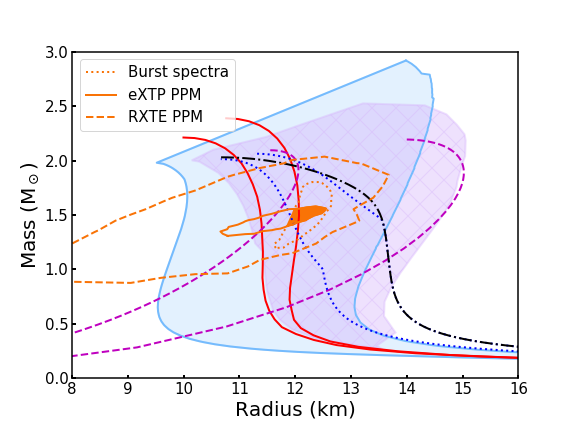}
\caption{Constraints from pulse profile modelling (PPM) and burst spectral fitting expected for eXTP.   EOS models as in Figure \ref{fig:eostomr}.  The orange dashed contours show the 1$\sigma$ constraints obtained from the pulse profiles of the AMP SAX~J1808.4--3658 using RXTE data \citep{Poutanen03,Salmi18}.  The constraints expected from pulse profile modelling and polarization data using the LAD and PFA on eXTP, for the same source, from an observation $\sim 100$ ks, are shown by the solid orange contour.  Spectral evolution during PRE bursts (observed from this source) as determined using direct fits with atmosphere spectral models \citep{Nattila17} produces more perpendicular constraints on M and R, given by dotted orange curves.  Combining these methods constrains the mass and radius of the neutron star to lie within the overlapping region (filled orange), with errors of a few \% on both parameters.  Pulse profile modelling of burst oscillations from this source will provide an entirely independent constraint on M and R with a similar level of accuracy.  These techniques can then be applied to other known sources (including several AMPs that also have burst oscillations) to deliver the multiple measurements necessary to map the EOS.}
\label{fig:extp_eos}
\end{figure*}

\subsection{Accretion-powered millisecond pulsars}
\label{sec:AMPS}

Accretion-powered millisecond pulsars (AMPs) contain weakly magnetised NSs (with $B\sim 10^8$--$10^9$~G) accreting matter from a  typically rather small companion star \citep{Patruno12}.  We now know of 16, all transients that go into outbursts every few years.   NSs in these systems have been spun up by accretion up to millisecond periods. Close to the NS, the accreting matter follows the magnetic field lines hitting the surface close to the magnetic poles. The resulting shockwave heats the electrons to $\sim$30--60 keV producing X-ray radiation  by thermal Comptonization in a slab of Thomson optical depth of order unity \citep{Poutanen06}.  Rotation of the hotspot causes modulation of the observed flux with the pulsar phase because of the evolving solid angle subtended on the observer's sky, as well as of Doppler boosting.  As the observed pulsations indicate that the shock covers only a small part of the NS surface, the scattered radiation should be linearly polarized up to 20\%, depending on the pulse phase, the photon energy and the geometry of the system.  In addition to the emission from the shock, pulsating thermal emission from the heated NS surface is seen at lower energies. In the peaks of the outbursts, when the accretion rate is high, the pulse profiles are usually very stable and rather sinusoidal with a harmonic content growing towards higher energies as a result of stronger contribution of the Comptonized emission which has a more anisotropic emission pattern.  The pulse shape implies that only a single hotspot is seen, while the secondary pole is blocked by the accretion disk.  The pulse stability allows the collection of millions of photons under constant conditions. 

One of the challenges for modelling pulse profiles from AMPs is the absence of first-principles models
that predict the emission pattern from the shock. The angular dependence, therefore, has to be parametrized, based on models of radiation transfer in an optically thin slab of hot electrons.  Degeneracy between the number of parameters did not allow strong constraints on M and R using existing data from the {\it Rossi X-ray Timing Explorer} (RXTE)  (see \citealt{Poutanen03}, \citealt{Leahy08,Leahy09,Leahy11}).  The LAD on eXTP would allow the collection of many more photons, and significant improvement in the constraints on M and R.  Furthermore, in a 100 ks observation of a bright AMP such as SAX J1808.4--3654 or XTE J1751--305 the X-ray polarimeter onboard of eXTP can measure polarisation in 10 phase bins at the 3$\sigma$ level and thus determine the basic geometrical parameters such as spot colatitude  and observer inclination (Fig.~\ref{fig:amps_polarization}). This not only improves the constraints on M and R  (see solid orange contour in Fig.~\ref{fig:extp_eos}), but allows an independent check of the fitting procedure based on the pulse profile alone.

Observations with the LAD  of the PRE bursts from the AMPs and analysis of their spectral evolution in the cooling tail give independent M-R constraints (see for example \citealt{Suleimanov11}, \citealt{Poutanen14}, \citealt{Nattila16}, \citealt{Nattila17}).
Using the currently most accurate method to directly fit atmosphere spectral models to the data \citep{Nattila17}, one would be able to reduce the error in radius to just a few \%, allowing us to put strong constraints on the EOS of cold dense matter (see dotted orange contour in Fig.~\ref{fig:extp_eos}).

\subsection{Burst oscillation sources}

Hotspots that form during thermonuclear explosions on accreting NSs give rise to pulsations known as burst oscillations (\citealt{Strohmayer06b}, \citealt{Galloway08}). The mechanism responsible for burst oscillations remains unknown: flame spread, uneven cooling, or even surface modes may play a role \citep[see][for a review]{Watts12}.   However burst oscillation sources are particularly attractive for M-R measurement in that they are numerous (increasing the odds of sampling a range of masses), have a well-understood thermal spectrum (\citealt{Suleimanov11b}, \citealt{Miller13b}), and offer multiple opportunities for independent cross-checks using complementary constraints \citep{Bhattacharyya05b, Chang05, Lo13}, thereby reducing systematic errors. Detailed studies have shown that accuracies of a few \% in M and R can be obtained with 10$^6$ pulsed photons (\citealt{Lo13}, \citealt{Psaltis14b}, \citealt{Miller15}). In addition the technique is robust, with clear flags if any of the assumptions made during the fitting process are breached. 

To estimate the observing time that eXTP would require to obtain measurements of M, R at the few \% level for known sources we can scale from the burst fluxes, burst oscillation amplitudes, burst recurrence times and the percentage of bursts with oscillations observed by RXTE.  For the persistent burst oscillation sources 4U 1636--536 and 4U 1728--34 we would require 350 ks and 375 ks respectively.  For burst oscillations from the transient AMPs SAX J1808.4--3658 and XTE J1814--338 we would require 490 ks and 275 ks respectively.   These observing times are substantial, but feasible.   Burst oscillations from AMPs are particularly useful since the M-R measurements they generate can be compared to the results obtained from pulse profile fitting of accretion powered pulsations from the same sources  (Sect.~\ref{sec:AMPS}).  In addition the constraints on system geometry (inclination) acquired from the phase-dependence of the polarization of the persistent emission can also be used in fitting the burst oscillations, reducing uncertainties on M and R.  Additional constraints for burst oscillation sources will also come from spectral fitting of strong bursts showing PRE (see Sect.~\ref{sec:AMPS} and \citealt{WP_OS}).

\begin{figure*}[!ht]
\includegraphics[width=0.9\textwidth]{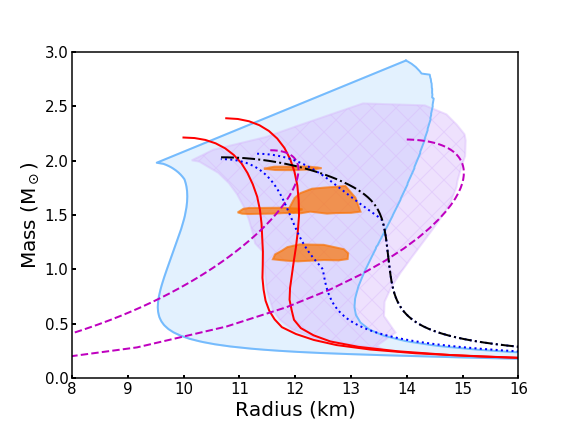}
\caption{Constraints from pulse profile modelling of rotation-powered pulsars with eXTP.   The orange error contours are for four MSPs for which masses are known precisely:   PSR J1614$-$2230,  PSR J2222$-$0137, PSR J0751$+$1807 and PSR J1909$-$3744.  The 1$\sigma$ contours shown correspond to $\sim$ 5--10\% constraints on R with the extent in M corresponding to the 1$\sigma$ bounds from the radio measurements.  This should be achievable using eXTP exposures of $\sim$ 1 Ms for each source.  The most stringent constraints on the EOS are likely to come from the highest mass target, and here more exposure time would be merited.   EOS models as in Figure \ref{fig:eostomr} -- the underlying model assumed in the simulations is the AP3 nucleonic EOS (red).}
\label{fig:rpp}
\end{figure*}

\subsection{Rotation-powered pulsars}
\label{rpp}

\textit{NICER} is a NASA Explorer Mission of Opportunity carrying soft X-ray timing instrument \citep{Arzoumanian14} that was installed on the International Space Station in June of 2017. \textit{NICER} applies the pulse profile modeling technique to X-ray emitting rotation-powered millisecond pulsars (MSPs) \citep{Bogdanov08}. Since \textit{NICER}'s targets rotate relatively slowly ($\sim$200 Hz), the measurements cannot rely on well-understood Doppler effects to break degeneracies between M and R. Nevertheless, if the surface radiation field and mass of the neutron star are known \textit{a priori}, \textit{NICER}  could in principle achieve an accuracy of $\sim$2\% in R (\citealt{Gendreau12}, \citealt{Bogdanov13}). The mass is now known to 5\% accuracy for \textit{NICER}'s main target, PSR J0437$-$4715 \citep{Reardon16}, but is not yet known for its other top targets. The surface radiation field depends on the pulsar mechanism and is at present not well constrained, although theoretical work to address this topic is underway.

\textit{NICER}  has a peak effective area at 1 keV of 1800 cm$^2$. \textit{eXTP} will be a factor of 4--5 larger in the soft waveband, enabling it to measure energy-resolved pulse waveforms of the nearest pulsars such as PSR J0437$-$4715 \citep{Bogdanov13} and J0030$+$0451 \citep{Bogdanov09} more efficiently than \textit{NICER}, thus producing improved constraints on M-R. Perhaps more importantly, the larger collecting area and significantly lower background of the \textit{eXTP}  SFA will enable studies of fainter MSPs that are not accessible with \textit{NICER}. Of great interest are nearby MSP binaries with precise measurements of the NS mass from radio pulse timing. These include PSR J1614$-$2230 with M$=1.928\pm0.017$ M$_{\odot}$ \citep{Fonseca16}, PSR J2222$-$0137  \citep[M$=1.20\pm0.14$ M$_{\odot}$; ][]{Kaplan14}, PSR J0751$+$1807  \citep[M$=1.64\pm0.15$ M$_{\odot}$; ][]{Desvignes16}, PSR J1909$-$3744   \citep[M$=1.54\pm0.027$ M$_{\odot}$;][]{Desvignes16}. The broad range of masses spanned by these systems is particularly beneficial for mapping out the dependence of R on M.  Figure \ref{fig:rpp} shows the level of constraints achievable within $\sim$ 1 Ms exposure times with eXTP for these sources.

\section{Spin measurement}
\label{Spinmeas}

\begin{figure*}[!ht]
\centering
\includegraphics[width=0.8\textwidth]{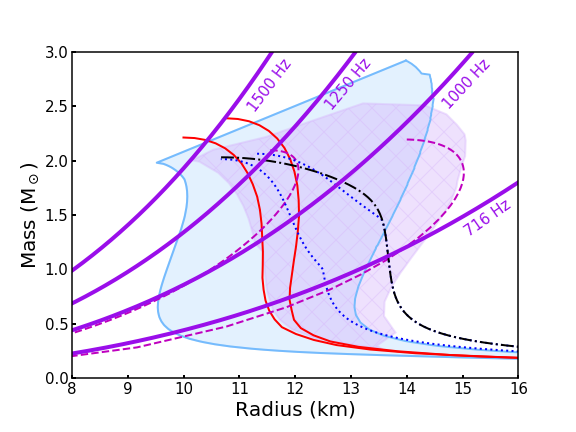}
\caption{Spin limits on the EOS.  The mass-shedding limit can be recast as an upper limit on radius for a star of a given spin rate \citep{Haensel09}.  This means that NSs of a given spin rate must extend to the left of the relevant limit in the M-R plane (shown as thick violet lines, for various spins).  EOS models as in Figure \ref{fig:eostomr}.  The current record holder, which spins at 716 Hz \citep{Hessels06} is not constraining. However, given a high enough spin individual EOS can be ruled out. Above 1000 Hz, for example, some individual EOS in the pQCD band, and one of the quark star models, would be excluded. }
\label{fig:spincons}
\end{figure*}

NSs with the fastest spins constrain the EOS since the limiting spin rate, at which the equatorial surface velocity is comparable to the local orbital velocity and mass-shedding occurs, is a function of M and R (Figure \ref{fig:spincons}). Softer EOS have smaller R for a given M, and hence have higher limiting spin rates.  More rapidly spinning NSs place increasingly stringent constraints on the EOS.  The current record holder (the MSP PSR J1748--2446ad in the Globular Cluster Terzan 5), which spins at 716 Hz \citep{Hessels06}, does not rotate rapidly enough to rule out any EOS models.  However the discovery of a NS with a sub-millisecond spin period would place a strong and clean constraint on the EOS.  There are prospects for finding more rapidly spinning NSs in future radio surveys \citep{Watts15}, however since the standard formation route for the MSPs is via spin-up due to accretion (\citealt{Alpar82}, \citealt{Radhakrishnan82}, \citealt{Bhattacharya91}), it is clear that we should look in the X-ray as well as the radio, and theory has long suggested that accretion could spin stars up close to the break-up limit \citep{Cook94b}.  Interestingly the drop-off in spin distribution at high spin rates seen in the MSP sample is not seen in the current (albeit much more limited) sample of accreting NSs \citep{Watts16}.   

Since eXTP would have a larger effective area than all preceding X-ray timing missions \citep[see][for a comparison]{WPinstrumentation}, it is well suited to discover many more NS spins, using both burst oscillations and accretion-powered pulsations.  We know from RXTE that the latter can be highly intermittent (\citealt{Galloway07}, \citealt{Casella08}, \citealt{Altamirano08}), perhaps due to the way that accretion flows are channeled onto weakly magnetized NSs \citep{Romanova08}, or because these systems are close to alignment \citep{Lamb09b}.  In addition, weak persistent pulsations are expected in systems where magnetic field evolution as accretion progresses has driven the system towards alignment \citep{Ruderman91}.  Searches for weak pulsations can exploit the sophisticated semi-coherent techniques being used for the \textit{Fermi} pulsar surveys (\citealt{Atwood06}, \citealt{Abdo09}, \citealt{Messenger11}, \citealt{Pletsch12}), which compensate for orbital Doppler smearing.  

eXTP will be able to detect burst oscillations in individual Type I X-ray bursts to amplitudes of 0.4\% (1.3\%) rms in the burst tail (rise) assuming a 10s (1s) integration time; by stacking bursts, sensitivity improves.  In estimating detectability of accretion-powered pulsations with eXTP we consider three source classes:  bright (e.g. Sco X-1), moderate (e.g. Aql X-1) and faint (e.g. XTE J1807--294)\footnote{The assumed fluxes are as follows: Sco X-1, 0.5-10 keV flux of $1.6 \times 10^{-7}$ \flux \citep{Dai07}; Aql X-1, 0.5-10 keV flux of $3.3 \times 10^{-9}$ \flux  \citep{Raichur11}; XTE J1807-294, 0.5-10 keV flux of $1.7 \times 10^{-10}$ \flux \citep{Campana03}.}.  We consider both coherent and semi-coherent searches.  Coherent searches consider a simple FFT in a short data segment so that we do not lose coherence of the signal as a
consequence of Doppler shifts induced by the orbital motion. We consider a duration of 128s, comparable to the duration of intermittent pulsation episodes seen in Aql X-1 \citep{Casella08}.  Under these assumptions, eXTP will be able to perform a coherent search for intermittent pulsations down to amplitudes of 0.04\% (bright), 0.3\% rms (moderate), 1.9\% rms (faint) (5 $\sigma$ single trial limits). 

For semi-coherent searches, we assume a 10 ks long observation, which need not be continuous, and coherence lengths (the segment over which we can search for individual trains of coherent pulsations) of 256~s and 512~s.  These assumptions are extremely conservative, and we would expect to be able to do better than this for many of our target sources, for which we know orbital parameters, reducing the number of templates to be searched.  For this type of search eXTP would be sensitive down to amplitudes of 0.01\% rms (bright), 0.1\% rms (moderate), and 0.6\% rms (faint) (5 $\sigma$ single trial limits).    

eXTP can also conduct blind searches of nearby (less than 3~kpc) \textit{Fermi} LAT sources that are suspected eclipsing redback and transitional MSP binaries similar to the canonical ``missing link'' PSR J1023+0038 \citep{Archibald15} and XSS J2124--3358 \citep{Bassa14}.  There are a handful of candidates that seem to be undetectable even in deep radio pulsation searches, but are by all other accounts strong redback MSP candidates.  The 716 Hz MSP in Terzan 5 is actually one of these eclipsing redback binaries, so conceivably some of these \textit{Fermi} sources may be harboring even faster MSPs.

\section{Constraints from accretion flows in the disks of NS Low Mass X-ray Binaries}

The advanced timing and polarimetry capabilities of eXTP will also enable other methods that could constrain the EOS for accreting NSs.  The methods outlined in this section are derived from phenomena associated with the inner parts of the accretion disk.  Compared to the spin rate constraint described in Section~\ref{Spinmeas} they are more model-dependent.  However, they are nonetheless powerful as they provide additional complementary cross-checks and allow us to calibrate different techniques to extend our reach to a wider range of sources.  See the eXTP White Papers on Strong Gravity \citep{WP_SG} and Observatory Science \citep{WP_OS} for further discussion of accretion flows.

\subsection{Kilohertz Quasi-Periodic Oscillations (QPOs)}

Kilohertz QPOs are rapid variations in the intensity of NS Low Mass X-ray Binaries (LMXBs), both persistent and transient \citep[see][for a review]{vanderKlis00}.  RXTE observed this phenomenon in a few tens of sources.   The corresponding millisecond time scale is so short that the QPOs must be associated with dynamical time scales in the accretion flow in the vicinity of NSs.   In many cases, these QPOs are seen as twin peaks in the Fourier power spectra. If one of the twin peaks is an indicator of the orbital motion in the accretion flow, it would put a constraint on NS mass and radius: the stable orbit must be outside the NS so at its smallest at either the NS radius or the innermost stable circular orbit (ISCO) \citep{Miller98b}.  

In addition to the association of the kHz QPOs with orbital motion in the innermost accretion flow onto NSs based on the millisecond time scales, there is increasing observational evidence that the kHz QPOs do indeed indicate the orbital frequency in the accretion flow (or boundary layer) surrounding the NS.  The frequency of the lower kHz QPOs is anti-correlated with the mHz QPO flux in 4U 1608--52, which is consistent with a modulation of the orbital frequency under radiation force from the NS \citep{Yu02}. The pulse amplitude changes significantly when the upper kHz QPO passes the spin frequency in the accretion-powered millisecond pulsar SAX J1808.4--3654, strongly suggesting that the QPO is produced by azimuthal motion at the inner edge of the accretion disk, most likely orbital motion \citep{Bult15}.    

The behaviour of the QPOs as they approach their highest frequencies was difficult to resolve with RXTE as both amplitude and coherence drop at this point, although the behaviour is consistent with that expected near the ISCO \citep{Barret06}.  eXTP will make breakthroughs by being able to track the QPOs to higher frequencies where the amplitudes are weaker, and to investigate QPO variability on timescales a factor $\sim 10$ shorter.  The latter is very important: QPOs in Sco X-1, for example, have been observed to drift by more than 22 Hz in 0.08~s \citep{Yu01}. 

The QPO coherence drop and rapid frequency drifts may be due to radiation force effects on the orbital frequency in the accretion flow, since an anti-correlation between kHz QPO frequency and X-ray flux was detected on the time scales of lower frequency QPOs (where the flux probably originates from the NS, see \citealt{Yu01}, \citealt{Yu02}). In sources with the most detections of kHz QPOs such as 4U\,1636--536, the maximum QPO frequency seems to be anti-correlated with the X-ray flux \citep{Barret05}. Both this anti-correlation and the QPO coherence variation can be explained by radiation force effects.  The rate at which the QPO frequencies change as a function of the QPO frequencies themselves also supports a scenario in which the inner part of the accretion disc is truncated at a radius that is set by the combined effect of viscosity and radiation drag \citep{Sanna12}.  This in turn can put constraints on the NS EOS by measurements of the maximum kHz QPO frequency and the X-ray flux \citep{Yu08}, although relativistic magnetohydrodynamical simulations with radiation will be needed to create models of sufficient accuracy. 

The energy-dependent time lags of the kHz QPOs \citep{deAvellar13} offer an independent constraint on the physical size of the accretion disc, and hence the NS. Together with the time-averaged spectrum of the source, a combination of the frequency, amplitude and time lag of these variability features over very short time scales (see for example \citealt{Lee01}, \citealt{Zhang17}, \citealt{Ribeiro17}) will provide the transfer function of the system.  This depends upon the physical size of the accretion disc and the corona, and hence can be used to further constrain the radius of the NS.  With eXTP, the maximum kHz QPO frequency measured in bright sources on short time scales, and in sources at lower flux levels, would increase by about 50 Hz (or 5\%).  Using the ISCO model of \citet{Miller98b}, this would lower the upper limit on the NS radius by $\sim$0.5 km or the mass by $\sim 0.1$ M$_\odot$ ($\sim 5$\% and $\sim$ 7\% respectively for a 10 km 1.4M$_\odot$ NS).  Corrections for radiation force effects would modify these estimates somewhat.

\begin{figure*}
\centering
\includegraphics[width=0.8\textwidth]{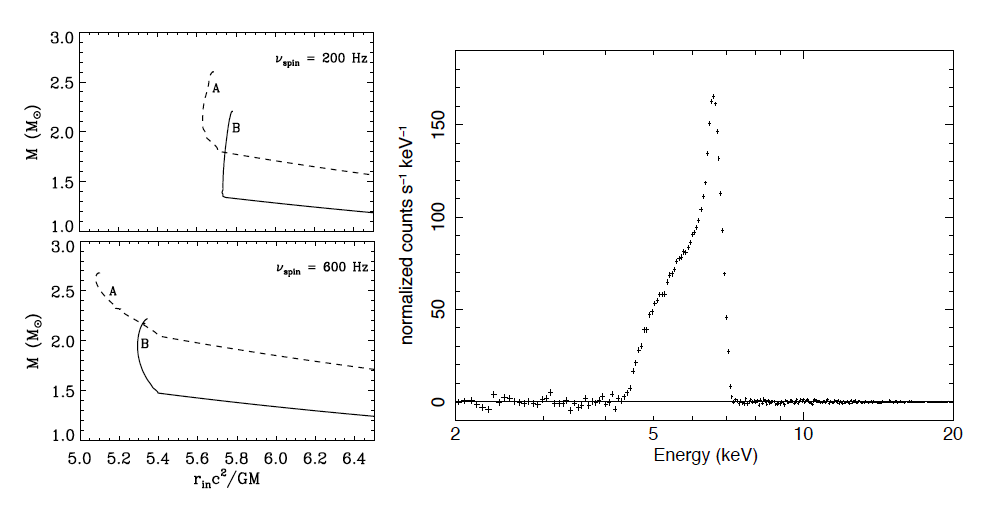}
\caption{EOS constraints from relativistic Fe line modelling.  Left panel \citep[adapted from][]{Bhattacharyya11}: M versus $r_{\rm in}c^2$/GM curves for two reasonable values of spin, and for two currently viable EOS models: one very stiff (A) and one of intermediate stiffness (B).  On the straight (near horizontal) portions, $r_{\rm in}$=R; elsewhere $r_{\rm in}=r_{\rm ISCO}$.   Since the geometrically thin accretion disk, which gives rise to the broad relativistic line, may be truncated by radiation pressure or a stellar magnetic field at a radius larger than $r_{\rm ISCO}$ and R, the observationally inferred $r_{\rm in}c^2$/GM is an upper limit. If M can be measured independently (e.g. by one of the other methods described in this white paper), this upper limit on $r_{\rm in}c^2$/GM provides an extra constraint on the EOS. In addition, if the upper limit of $r_{\rm in}c^2$/GM is sufficiently small, softer EOS models can be ruled out without a mass measurement.  Right:  Simulated relativistic Fe line spectrum for a 30ks observation with eXTP (residual plot).  We assume a {\tt wabs(bbody+diskbb+powerlaw+diskline)} XSPEC model with reasonable values of 2--20~keV flux ($6.4\times10^{-9}$ erg cm$^{-2}$ s$^{-1}$) and line equivalent width (124 eV).}
\label{fig:Fefig}
\end{figure*}

\subsection{Constraints from relativistic Fe line modelling}

A broad relativistic Fe K$\alpha$ spectral emission line is observed from many stellar-mass and supermassive black hole systems (\citealt{Fabian00}, \citealt{Reynolds03}, \citealt{MillerJ07}). Such a fluorescent line near 6 keV is believed to be generated by the reflection of hard X-rays from the accretion disk, and is shaped by various physical effects, such as the Doppler effect, special relativistic beaming, gravitational redshifting and GR light-bending. The properties of this line can be used to measure $r_{\rm in}c^2$/GM, i.e., the inner-edge radius $r_{\rm in}$ of the accretion disk in the unit of the black hole mass M. By considering the disk inner-edge to be the innermost stable circular orbit (ISCO), which may be a reasonable assumption for black holes, one can also infer the black hole angular momentum parameter for the Kerr spacetime.

A broad relativistic spectral line has also been observed  from a number of NS LMXBs (\citealt{Bhattacharyya07}, \citealt{Cackett08}, \citealt{Pandel08}, \citealt{Dai09}, \citealt{Cackett10}, \citealt{MillerJ13}, \citealt{Chiang16}).  As for black holes, one can infer $r_{\rm in}c^2$/GM for NSs from the relativistic Fe line.  Since the disk inner edge radius $r_{\rm in} \ge$ R, the inferred $r_{\rm in}c^2$/GM provides an upper limit on R$c^2$/GM. One can therefore use M-$r_{\rm in}c^2$/GM space (instead of M-R space) for a known spin to constrain EOS models (Figure \ref{fig:Fefig} and \citealt{Bhattacharyya11}). This method requires computations of $r_{\rm in}c^2$/GM for given M, spin and EOS models. Note that, while $r_{\rm in} = r_{\rm ISCO}$ (i.e., ISCO radius) for a black hole,  $r_{\rm in}$ is either $r_{\rm ISCO}$ or R, whichever is greater, for a NS. For a spinning (Kerr) black hole, $r_{\rm ISCO}$ can be analytically computed as a function of M and $a$. For a NS in an LMXB, one must compute $r_{\rm ISCO}$ and R values numerically for various EOS models and NS configurations, using an appropriate rapidly spinning stellar spacetime.  Simulations for the eXTP LAD show that a statistical error of less than 0.1 in $r_{\rm in}c^2/GM$, sufficient to distinguish models, is achievable with a 30 ks exposure (see \citealt{Bhattacharyya17} and Figure \ref{fig:Fefig}).

\section{Summary}

eXTP offers unprecedented discovery space for the EOS of cold supranuclear density matter.  eXTP's large area will enable the most sensitive searches for accretion-powered pulsations and burst oscillations ever undertaken.  Both yield the spin frequency of the NS; a single measurement of sub millisecond period spin would provide a clean and extremely robust constraint on the EOS.  

However, eXTP will also deliver high precision measurements of M and R. The combination of large effective area and polarimeter will enable us to deploy multiple independent techniques:  pulse profile modelling of accretion-powered pulsations, burst oscillations, and rotation-powered pulsations; spectral modelling of bursts, and using phenomena related to the accretion disc such as kHz QPOs and the relativistic Fe line.  Many sources show several of these phenomena, allowing us to make completely independent measurements for a single source, to reduce systematic errors.   Examples of targets in this class include the accretion-powered millisecond pulsar SAX J1808.4--3658, which goes into regular outburst, and the persistently accreting burster 4U 1636--536.  We anticipate that eXTP could delivery precision constraints on M and R, at the few percent level, for of order 10 sources for a reasonable observing plan and given the anticipated mission lifetime.    This would be unprecedented in terms of mapping the EOS and expanding the frontiers of dense matter physics.

{\bf Acknowledgments}: ALW and TER acknowledge support from ERC Starting Grant 639217 CSINEUTRONSTAR.  AP acknowledges support from a Netherlands Organization for Scientific Research (NWO) Vidi Fellowship.  YC is suported by the European Union’s Horizon 2020 research and innovation programme under the Marie Sklodowska-Curie Global Fellowship grant agreement No 703916.  SKG, KH and AS are supported in part by the DFG through Grant SFB 1245 and the ERC Grant No. 307986 STRONGINT.

{\bf Author contributions}:  This paper is an initiative of eXTP's Science Working Group 1 on Dense Matter, whose members are representatives of the astronomical community at large with a scientific interest in pursuing the successful implementation of eXTP. The paper was primarily written by Anna Watts, Wenfei Yu, Juri Poutanen, and Shu Zhang with major contributions by Sudip Bhattacharyya (Fe lines), Slavko Bogdanov (rotation powered pulsars), Long Ji (spin measurements), Alessandro Patruno (spin measurements) and Thomas Riley (pulse profile modelling technique).  Contributions were edited by Anna Watts.  Other co-authors provided input to refine the paper.

\end{multicols}
\end{document}